\documentclass{PoS}

\usepackage{xspace}
\usepackage{subfig}
\usepackage{axodraw4j}
\usepackage{mcite}

\newcommand{\Sherpa}{S\protect\scalebox{0.8}{HERPA}\xspace}
\newcommand{\sbr}[1]{\left[ #1\right]}
\newcommand{\kinsketches}{
  \subfloat[][]{\scalebox{0.33}{
  \begin{picture}(328,133) (303,-159)
    \SetWidth{1.0}
    \SetColor{Black}
    \Line[arrow,arrowpos=0.5,arrowlength=5,arrowwidth=2,arrowinset=0.2](352,-91)(416,-43)
    \Line[arrow,arrowpos=0.5,arrowlength=5,arrowwidth=2,arrowinset=0.2](416,-123)(352,-91)
    \Line(629,-116)(581,-116)
    \Line(629,-97)(581,-97)
    \Line[arrow,arrowpos=0.5,arrowlength=5,arrowwidth=2,arrowinset=0.2](630,-107)(582,-107)
    \GOval(582,-107)(22,6)(0){0.882}
    \Line[arrow,arrowpos=0.5,arrowlength=5,arrowwidth=2,arrowinset=0.2](576,-107)(496,-107)
    \Line[arrow,arrowpos=0.5,arrowlength=5,arrowwidth=2,arrowinset=0.2](304,-155)(352,-91)
    \Line[arrow,arrowpos=0.5,arrowlength=5,arrowwidth=2,arrowinset=0.2](352,-91)(304,-27)
    \GOval(352,-91)(16,16)(0){0.6}
    \Line[arrow,arrowpos=0.5,arrowlength=5,arrowwidth=2,arrowinset=0.2](496,-107)(416,-123)
    \Gluon(464,-75)(496,-107){7.5}{4}
    \Gluon(416,-123)(384,-155){7.5}{4}
  \end{picture}}\label{fig:multicore_one}}\hspace{1cm}
  \subfloat[][]{\scalebox{0.33}{
  \begin{picture}(328,163) (303,-142)
    \SetWidth{1.0}
    \SetColor{Black}
    \Line[arrow,arrowpos=0.5,arrowlength=5,arrowwidth=2,arrowinset=0.2](496,-93)(416,-61)
    \Line[arrow,arrowpos=0.5,arrowlength=5,arrowwidth=2,arrowinset=0.2](416,-61)(400,-141)
    \Line[arrow,arrowpos=0.5,arrowlength=5,arrowwidth=2,arrowinset=0.2](576,-93)(496,-93)
    \Line[arrow,arrowpos=0.5,arrowlength=5,arrowwidth=2,arrowinset=0.2](304,-125)(352,-61)
    \Line[arrow,arrowpos=0.5,arrowlength=5,arrowwidth=2,arrowinset=0.2](352,-61)(304,3)
    \GOval(416,-61)(16,16)(0){0.6}
    \Gluon(496,-93)(464,-125){7.5}{4}
    \Photon(352,-61)(400,-61){5}{3}
    \Line(629,-102)(581,-102)
    \Line(629,-83)(581,-83)
    \Line[arrow,arrowpos=0.5,arrowlength=5,arrowwidth=2,arrowinset=0.2](630,-93)(582,-93)
    \GOval(582,-93)(22,6)(0){0.882}
    \Gluon(416,-45)(448,19){7.5}{6}
  \end{picture}}\label{fig:multicore_two}}\hspace{1cm}
  \subfloat[][]{\scalebox{0.33}{
  \begin{picture}(328,163) (303,-159)
    \SetWidth{1.0}
    \SetColor{Black}
    \Line[arrow,arrowpos=0.5,arrowlength=5,arrowwidth=2,arrowinset=0.2](400,-61)(464,-45)
    \Line[arrow,arrowpos=0.5,arrowlength=5,arrowwidth=2,arrowinset=0.2](480,-77)(400,-61)
    \Line[arrow,arrowpos=0.5,arrowlength=5,arrowwidth=2,arrowinset=0.2](304,-125)(352,-61)
    \Line[arrow,arrowpos=0.5,arrowlength=5,arrowwidth=2,arrowinset=0.2](352,-61)(304,3)
    \GOval(496,-77)(16,16)(0){0.6}
    \Line[arrow,arrowpos=0.5,arrowlength=5,arrowwidth=2,arrowinset=0.2](576,-77)(512,-77)
    \Gluon(496,3)(496,-61){7.5}{5}
    \Photon(352,-61)(400,-61){5}{3}
    \Line(629,-86)(581,-86)
    \Line(629,-67)(581,-67)
    \Line[arrow,arrowpos=0.5,arrowlength=5,arrowwidth=2,arrowinset=0.2](630,-77)(582,-77)
    \GOval(582,-77)(22,6)(0){0.882}
    \Gluon(496,-93)(464,-157){7.5}{6}
  \end{picture}}\label{fig:multicore_three}}
}

\title{Hadronic final states in DIS with \Sherpa}
\ShortTitle{Hadronic final states in DIS with \Sherpa}
\author{Tancredi Carli\\
        CERN, Department of Physics, CH-1211 Geneva 23, Switzerland\\
        E-mail: \email{tancredi.carli@cern.ch}}
\author{Thomas Gehrmann\\
        Universit{\"a}t Z{\"u}rich, CH-8057 Z{\"u}rich, Switzerland\\
        E-mail: \email{thomas.gehrmann@physik.uzh.ch}}
\author{\speaker{Stefan H{\"o}che}\\
        Universit{\"a}t Z{\"u}rich, CH-8057 Z{\"u}rich, Switzerland\\
        E-mail: \email{shoeche@physik.uzh.ch}}

\abstract{We present an extension of the multi-purpose Monte-Carlo event generator 
  \Sherpa for processes in deeply inelastic lepton-nucleon scattering. Hadronic 
  final states in this kinematical setting are characterised by the presence 
  of multiple kinematical scales, which were up to now accounted for only 
  by specific resummations in individual kinematical regions. An extension 
  of a known method for merging truncated parton showers with higher-order 
  tree-level matrix elements allows to obtain predictions which are reliable
  in all kinematical limits.}

\FullConference{XVIII International Workshop on Deep-Inelastic Scattering and Related Subjects\\
  April 19 -23, 2010\\
  Convitto della Calza, Firenze, Italy}

\begin{document}

\section{Introduction}
Deep-inelastic lepton-nucleon scattering (DIS) offers the possibility to study
the structure of the nucleon and the dynamics of strong interactions by means
of a pointlike probe. It provides a clean experimental setting to analyse 
inclusive quantities like for example the proton structure functions. 
However, the kinematical configurations in DIS are usually very different 
from those in other processes in collider experiments. The virtuality of the exchanged
photon tends to be close to zero, while final state jets might still have 
large transverse momenta. Experiments conducted at the HERA collider have shown 
that the large available phase space can lead to a considerable number of hard jets, 
even if the photon virtuality is low.

This fact poses a certain problem for the simulation of DIS
with Monte-Carlo event generators. Such programs usually employ parton showers 
based on the DGLAP equations~\cite{Gribov:1972ri,*Lipatov:1974qm,
  *Dokshitzer:1977sg,*Altarelli:1977zs}. It is assumed that any scattering process 
factorises into a core $2\to 2$ interaction and a shower evolution, which simply
``dresses'' the hard interaction with softer radiation. As the only hard scale 
set by the leading-order DIS process $e^{\pm}q\to e^{\pm}q$ is the photon virtuality, 
$-Q^2$, the probability to produce a jet of transverse momentum larger than $Q^2$ 
would then vanish. In order to reliably simulate DIS events, one must therefore resort
to different techniques. We aim at an approach based on combining higher-order tree-level 
matrix elements with the parton shower along the lines of Ref.~\cite{Hoeche:2009rj},
which is consistent with the DGLAP framework. The technical prerequisites 
for realising this method are found in the multi-purpose Monte-Carlo event generator 
\Sherpa{}~\cite{Gleisberg:2003xi,*Gleisberg:2008ta}.

\section{Event generation technique}
The basic idea of the approach is to separate the phase space into a matrix-element 
and a parton-shower domain through a cut in the phase space of multi-parton processes. 
The matrix-element domain is then supposed to contain hard, well-separated partons only, 
while the parton-shower domain covers the region where resummation effects become 
important. Throughout the hard domain parton-shower emissions are corrected 
using tree-level matrix elements up to a given maximum multiplicity. In the soft domain, 
the parton shower is applied as is. The separation is achieved in terms of a so-called 
jet criterion, defining the ``hardness'' and/or the separation of a parton with respect 
to others~\cite{Hoeche:2009rj}. Several successful studies of various classes of processes 
have demonstrated the capability of this technique to correctly describe multi-jet
final states~\cite{Alwall:2007fs,*Mangano:2006rw,*Alwall:2008qv,Hoeche:2009xc}.
\begin{figure}
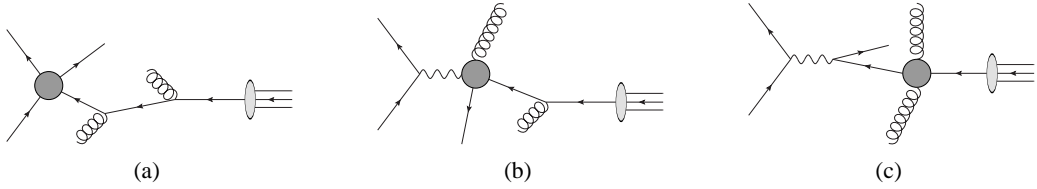

  \begin{center}
  \kinsketches
  \caption{Schematic view of three possible core process choices in DIS three-jet production. 
  Part~\protect\subref{fig:multicore_one} corresponds to the most probable core process
  being the virtual photon exchange, while additional hard partons are interpreted
  as parton shower emissions. Parts~\protect\subref{fig:multicore_two} 
  and~\protect\subref{fig:multicore_three} depict configurations, where the most probable
  core process is the interaction of the virtual photon with a parton and a pure QCD
  $2\to 2$ process, respectively.
  \label{fig:multicore}}
  \end{center}
\end{figure}

As pointed out in~\cite{Hoeche:2009rj}, the above merging algorithm needs to be refined
if the scale difference between $Q^2$ and the hardness scale $k_T^2$ of additional partons 
is large and negative. In this case, logarithmic corrections are not induced by $Q^2/q^2$, 
but rather by $k_T^2/q^2$, where $q^2$ is the jet resolution scale. The production of 
the virtual photon can then be regarded as an electroweak splitting process, attached to 
a core interaction of type $\gamma^* j\to jj$, as depicted in Fig.~\ref{fig:multicore_two}.
In the extreme case of very hard jets, the core process does not even include the 
virtual photon, cf.\ Fig.~\ref{fig:multicore_three}. The main task of the algorithm is 
to correctly identify the ``core'' interaction underlying a multi-parton process and 
to employ it to define starting conditions for the parton shower evolution~\cite{Carli:2009cg}.

Conversely, this idea can be used to lift the restriction on the real-emission phase space
at low $Q^2$. A similar method is in fact employed in Drell-Yan lepton-pair 
production via $\gamma^*/Z$-exchange, where the separation cut $Q_{\rm cut}$ between 
matrix-element and parton-shower domain is set such that $Q_{\rm cut}<m_{ll'}$, with $m_{ll'}$
being the invariant mass of the lepton pair. This choice implies that jets of $k_T^2\gtrsim m_{ll'}$
always fall into the matrix element domain. In deep-inelastic-scattering the situation is 
slightly different due to the variable value of $Q^2$, which plays the role of $m_{ll'}$ 
in the Drell-Yan pair production case. The solution can, however, be identical. We choose
\begin{equation}\label{eq:sliding_qcut}
  Q_{\rm cut}\,=\;\bar{Q}_{\rm cut}\,\sbr{\;1+\frac{\bar{Q}_{\rm cut}^2/S_{\,\rm DIS}^{\,2}}{Q^2}\;}^{-1/2}\;,
\end{equation}
where $\bar{Q}_{\rm cut}$ is a fixed value, much like $Q_{\rm cut}$ in the Drell-Yan pair production case
and $S_{\,\rm DIS}<1$ is a constant with lower limit enforced by experimental requirements.
$\bar{Q}_{\rm cut}$ ensures that high-$Q^2$, medium-$E_{T,B}^2$ events are described by matrix elements, rather 
than by the parton shower. At the same time, the factor in the square bracket enforces low-$Q^2$, 
high-$E_{T,B}^2$ events to be in the matrix-element domain as well, such that the complete available 
real-emission phase space can be filled.

\section{Comparison with experimental data}
In this section, we show some comparison with experimental data to exemplify the performance of
the Monte Carlo simulation. The correct description of the selected measurements is quite challenging 
for the Monte Carlo traditionally used in the analysis of HERA data~\cite{Brook:1995nn,*Brook:1998jd}.

\begin{figure}
  \begin{center}
  \subfloat[][]{\includegraphics[width=0.495\textwidth]{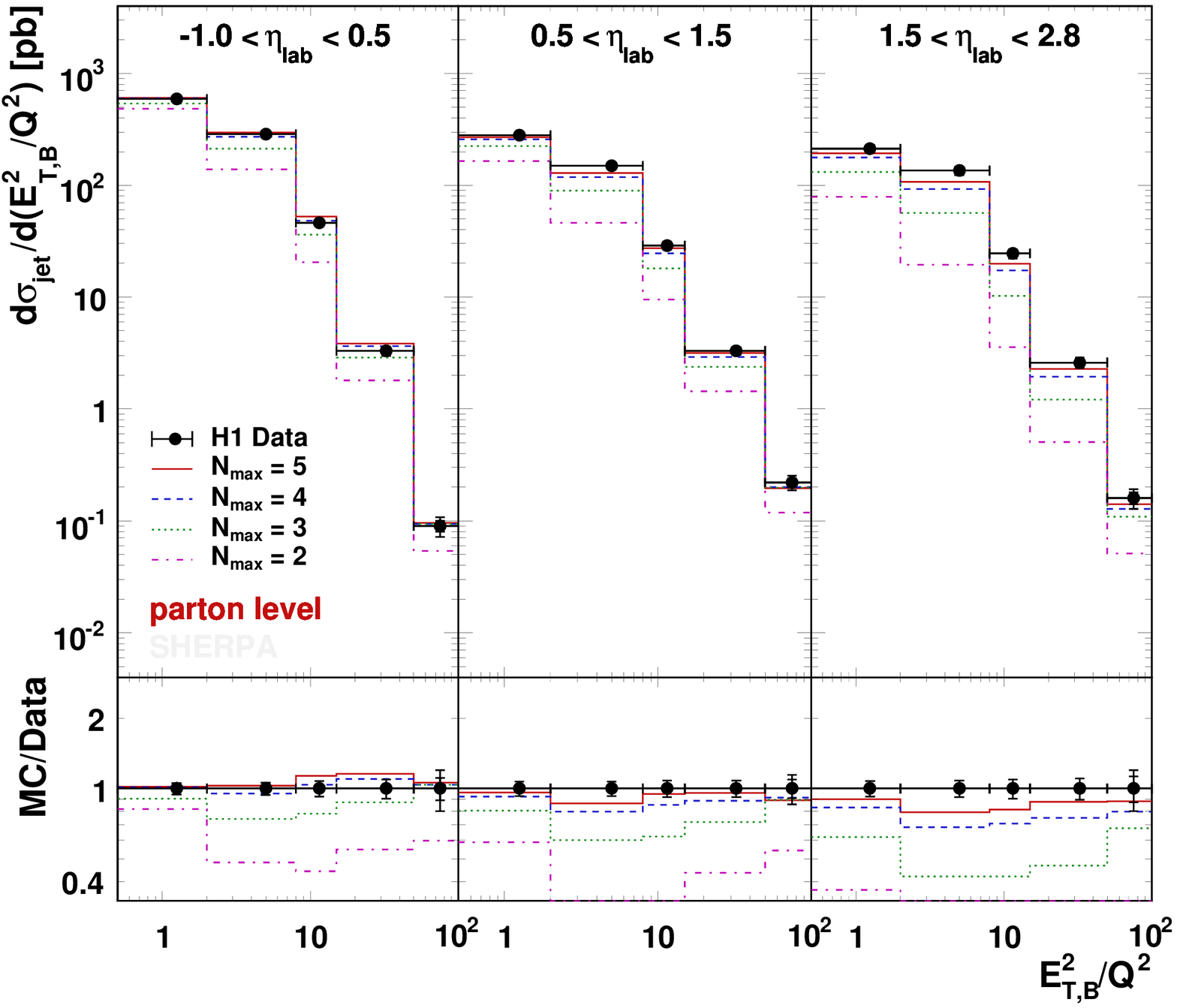}
    \label{fig:etq2_njet}}
  \subfloat[][]{\includegraphics[width=0.495\textwidth]{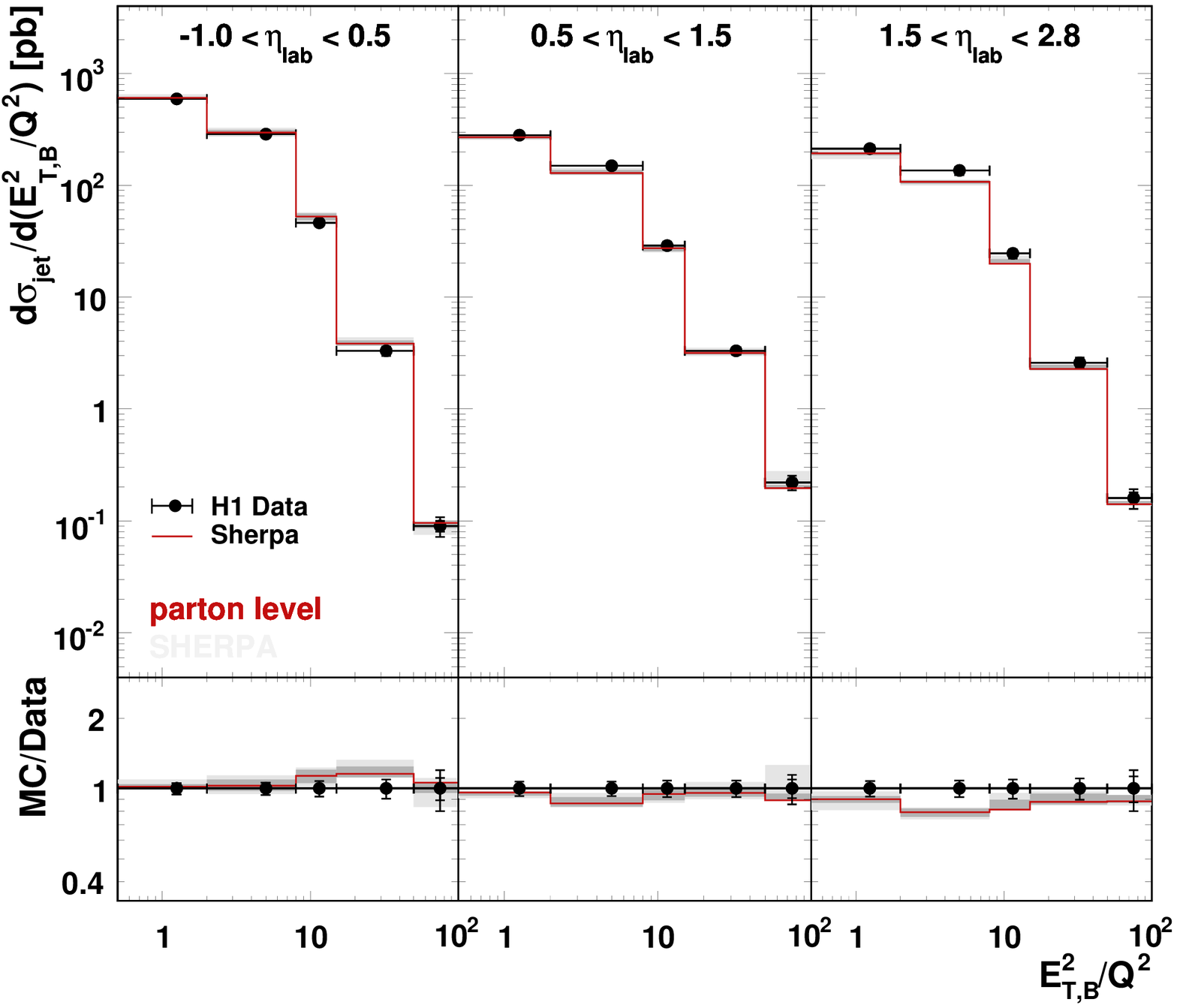}
    \label{fig:etq2_qcut}}
  \caption{The inclusive jet cross section as a function of $E_{T,B}^2/Q^2$ in bins
  of $\eta_{lab}$, measured by the H1 Collaboration~\protect\cite{Adloff:2002ew}.
  $E_{T,B}^2$ is the jet transverse energy in the Breit frame, while $\eta_{lab}$
  denotes the jet rapidity in the laboratory frame. 
  Part~\protect\subref{fig:etq2_njet} displays the influence of the maximum parton multiplicity,
  $N_{\,\rm max}$, from hard matrix elements. We show the uncertainty originating from varying
  $S_{\,\rm DIS}$ between 0.5 and 0.7 (light grey band) and from varying $\bar{Q}_{\rm cut}$ 
  between 3 GeV and 9 GeV (dark grey band) in part~\protect\subref{fig:etq2_qcut}.}
  \label{fig:etq2}
  \end{center}
\end{figure}

A crucial observable is given by the inclusive jet cross section, differential 
with respect to $E_{T,B}^2/Q^2$, where $E_{T,B}$ is the jet transverse energy 
in the Breit frame. For $E_{T,B}^2/Q^2>1$ it probes a part of the phase space where 
leading order Monte-Carlo models without the inclusion of low-$x$ effects are bound 
to fail in their description of jet spectra.
Figure~\ref{fig:etq2_njet} shows that the Monte-Carlo prediction gradually 
improves with a growing number of final-state partons in the hard matrix elements.
The uncertainties associated with a variation of the intrinsic parameters of the 
merging algorithm are shown in Fig.~\ref{fig:etq2_qcut}.

\begin{figure}
  \begin{center}
    \includegraphics[width=0.495\linewidth]{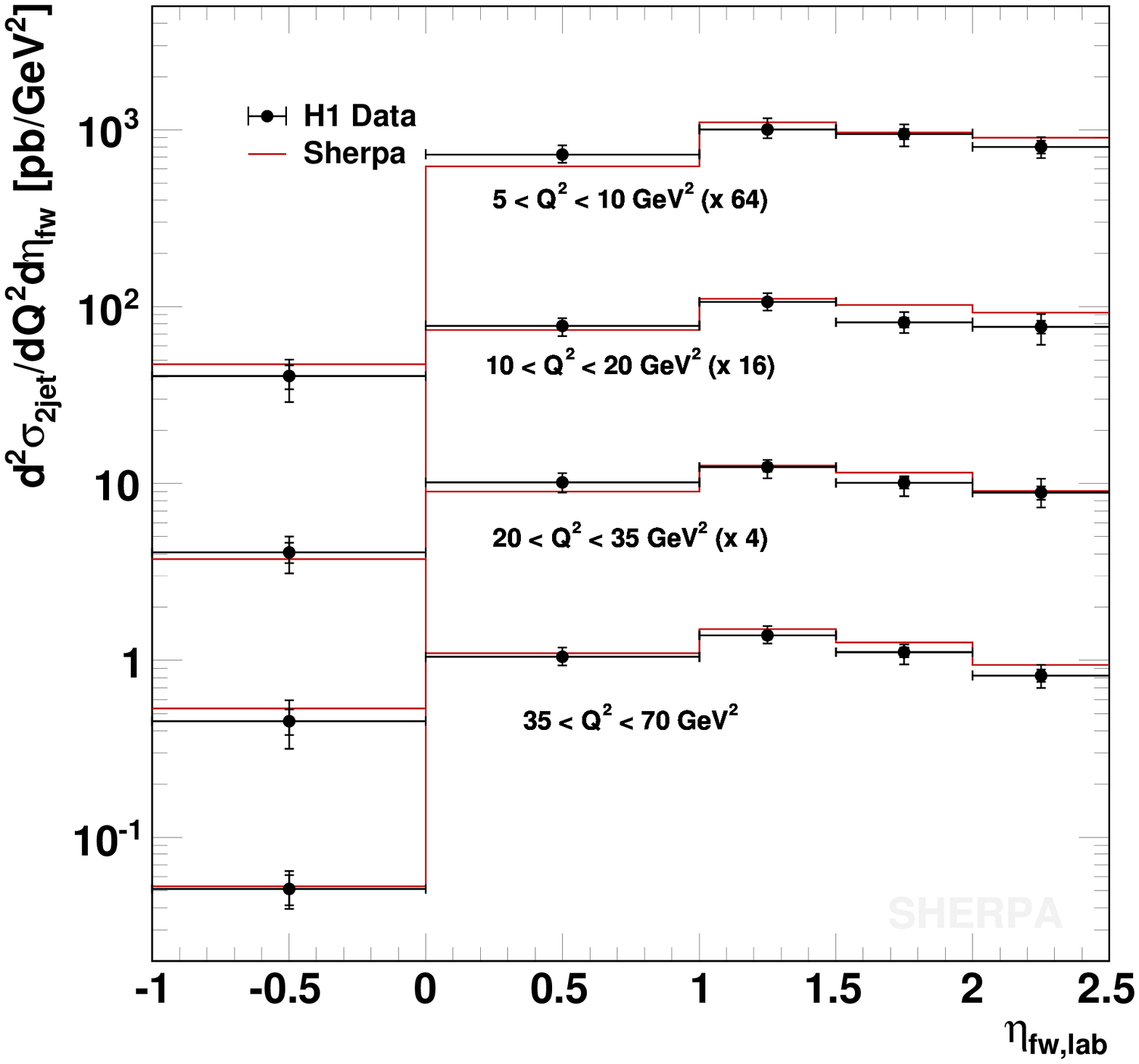}
    \includegraphics[width=0.495\linewidth]{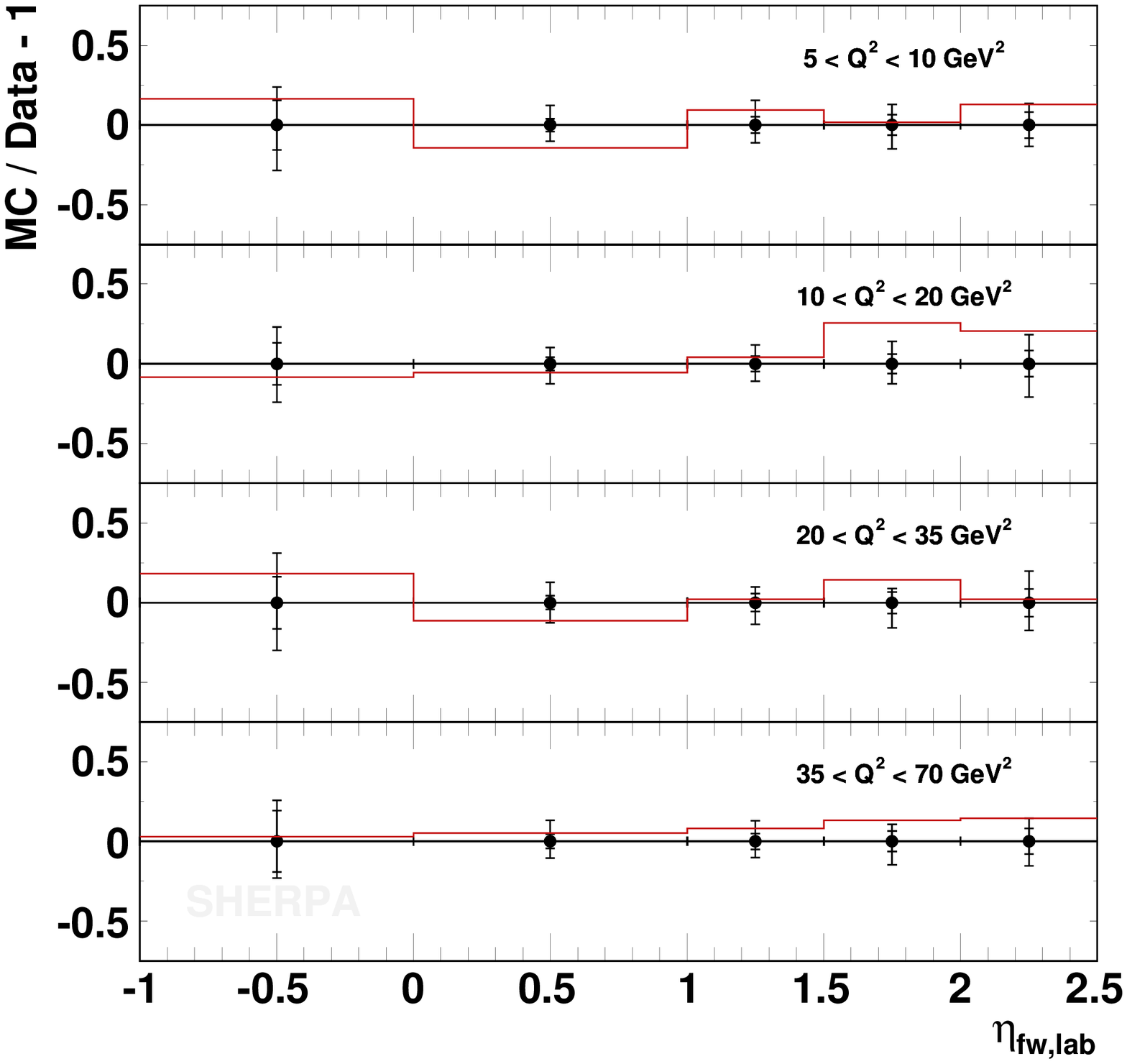}\\
    \includegraphics[width=0.495\linewidth]{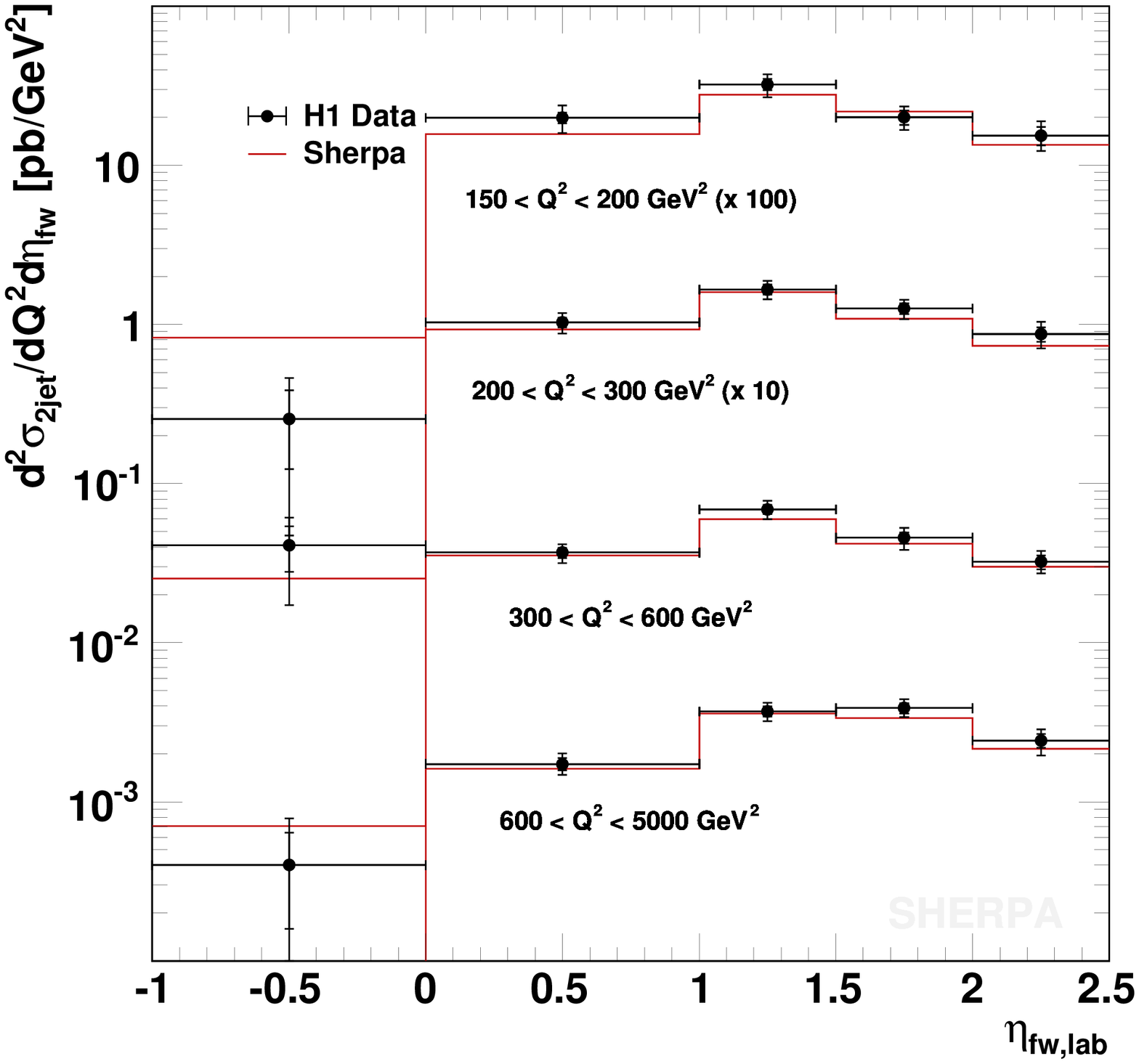}
    \includegraphics[width=0.495\linewidth]{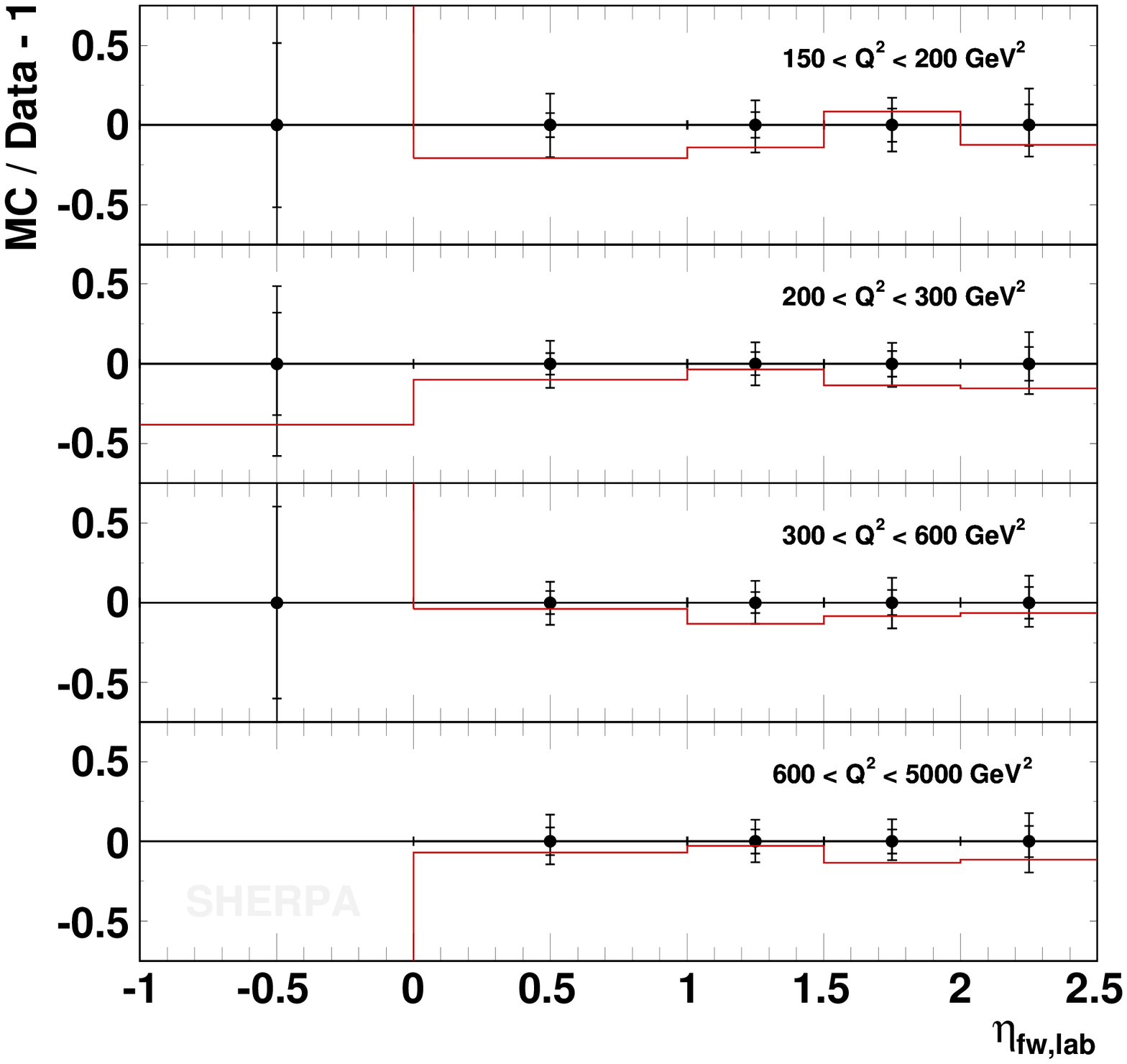}
  \end{center}
  \caption{The di-jet cross section as a function of $\eta_{\rm fw, lab}$ in bins 
  of $Q^2$, measured by the H1 Collaboration~\protect\cite{Adloff:2000tq}.
  \label{fig:eta_fw}}
\end{figure}

It is interesting to investigate jet properties in some more detail.
Figure~\ref{fig:eta_fw} displays rapidity spectra of the forward jet in di-jet production 
for various regions of $Q^2$. As for the case of inclusive jet production we observe 
a good description of the corresponding H1 data.

\section{Conclusions}
The \Sherpa event-generation framework has been extended to describe 
hadronic final states in deep-inelastic lepton-nucleon scattering processes. 
The simulation is based on merging higher-order tree-level matrix elements
with a parton shower. When applying this technique to DIS processes, it is
vital to correctly identify the core interaction, which can be either 
electron-quark scattering, photon-parton scattering or a partonic $2\to 2$ 
interaction, depending on the final state kinematics.
The particular kinematical situation in DIS also requires to choose 
appropriate merging scales, depending on the photon virtuality $Q^2$.
By doing so, we obtain a reliable description of DIS in all kinematical 
regions, including for example high-$E_{T,B}^2$, low-$Q^2$ processes.


\end{document}